\documentclass[aps]{revtex4}
\usepackage{amsmath}
\usepackage{epsfig, graphicx}
\begin{document}
 
\title{The Effect of Neutron Imaging Pinholes on Image Blurring}

\author{ Anemarie DeYoung and Anna Hayes}
\affiliation{Los Alamos National Laboratory, Los Alamos, NM  87545}

\date{\today}

\begin{abstract}

\hspace*{-1.3in}
\rotatebox{90}{%
\fbox{\parbox[t]{0.95in}{LA-UR-23-27192}}}
\vspace*{-1.05in}

We examine the effect of realistic pinhole geometries used in neutron imaging on the size of the image produced in the
image plane.
Using a simple inverted parabola shape for the neutron source, an MCNP simulation shows that the area of the images 
are only very slightly changed by pinhole blurring. 
Using a metric, $A80$, which  is the area of the image that contains 80\% of the neutrons, we find a maximum of 1.5\% increase in the metric.
In comparing a `perfect’ cylinder bore hole approximation for the pinhole with a full stepped collimator tapered 
pinhole we find a difference in $A80$ of 0.8\%.  
In addition, we find very little difference in the prediction of a full MCNP simulation compared 
to a simple ray-tracing calculation that includes neutron absorption in
the pinhole system but does not include neutron scattering.
We note that more sophisticated  assumptions for the neutron source shape than made here could lead to larger increases in the image size, and increases in $A80$ of the order of 5-8\% could be realistic.
\end{abstract}

\maketitle
\section{Introduction}
Neutron imaging of burning plasmas is an important diagnostic from systems ranging for inertial confinement fusion \cite
{Merrill, Verena} to testing of nuclear explosive devices using a diagnostic known as PINEX, which is 
short for pinhole camera experiment \cite{Sterling, King}.
In the case of the Underground Test (UGT) Program at the Nevada Test Site, neutron imaging techniques were used to measure neutron flux, energy and spatial distribution as a function of time \cite{Sterling, King}. 
Since it is difficult to ‘focus’ a neutron beam, or change the neutron angle in order to form an image with a lens, such as is done with light,  pinhole imaging was used instead. Neutron pinhole imaging relies on the neutrons following straight paths as in ray tracing, but blocks all neutrons outside of the small pinhole. In this way,  an inverted image is derived at the camera plane from the image in the plane of the neutron source. 

In order to block neutrons that don't pass through the pinhole from reaching the image plane, 
the pinhole must be manufactured from dense material. For this, a high-density alloy of tungsten, Kennertium, is often used and this has been shown to have
good radioactive shielding, high strength, room temperature ductility and good machinability\cite{kenn1}.
The mean free path (MFP) of 14.07 MeV neutrons in Kennertium is 3 cm. 
The geometrical design of the pinhole determines the field of view, the pinhole roll-off of the neutron intensity at the edges of the image, the blurring of the image from the so-called point spread function (PSF) and the overall neutron transmission. The PSF is the blurring of the image due to the pinhole geometry and is determined by calculating the image that would be measured for a point source. The observed image is then a convolution of the PSF with the original object,
\begin{equation}
S_I(x,y) = \int\int_{-\infty}^{+\infty} S_{M0}(x_0,y_0)p(x-x_0,y-y_0) dx_0 dy_0 +\eta(x,y)
\end{equation}
where $S_I(x,y)$ is the blurred image at the image plane, $S_{M0}(x_0,y_0)$ is the original image after magnification  and $p(x,y)$ is the PSF that defines the system degradation. 
$\eta(x,y)$ is background noise that is introduced by the system and the camera. 
If the PSF is known and the noise can be subtracted, the image can be largely de-blurred and the quality of the image restoration depends on the accuracy of PSF.
 
An ideal pinhole would be an infinitely thin and perfectly absorbing barrier \cite{Verena}. But the long mean free path of the neutrons means that the pinhole must be
thick and the best pinhole would then be a cylinder with a simple center bore, Fig. 1a.  The resultant PSF 
at the image plane is a top-hat shape with diameter $d_{top-hat}=d_{pinhole} \times (1 + M)$, where $d_{pinhole}$ is the diameter of the pinhole, and $M$ is the magnification of the system, $M=\ell_1/\ell_2$, where $\ell_1$ is the distance from the image to the pinhole and $\ell_2$ is the distance from the pinhole to the  source. 
However, there is a serious problem with a purely cylindrical pinhole because it leads to a small field of view, Fig. 1a. 
For this reason, tapers are used, meaning that the pinhole is a  straight cylinder with conical tapers on each side, Fig.1 b.
\begin{figure}
\includegraphics[width=4.0 in]{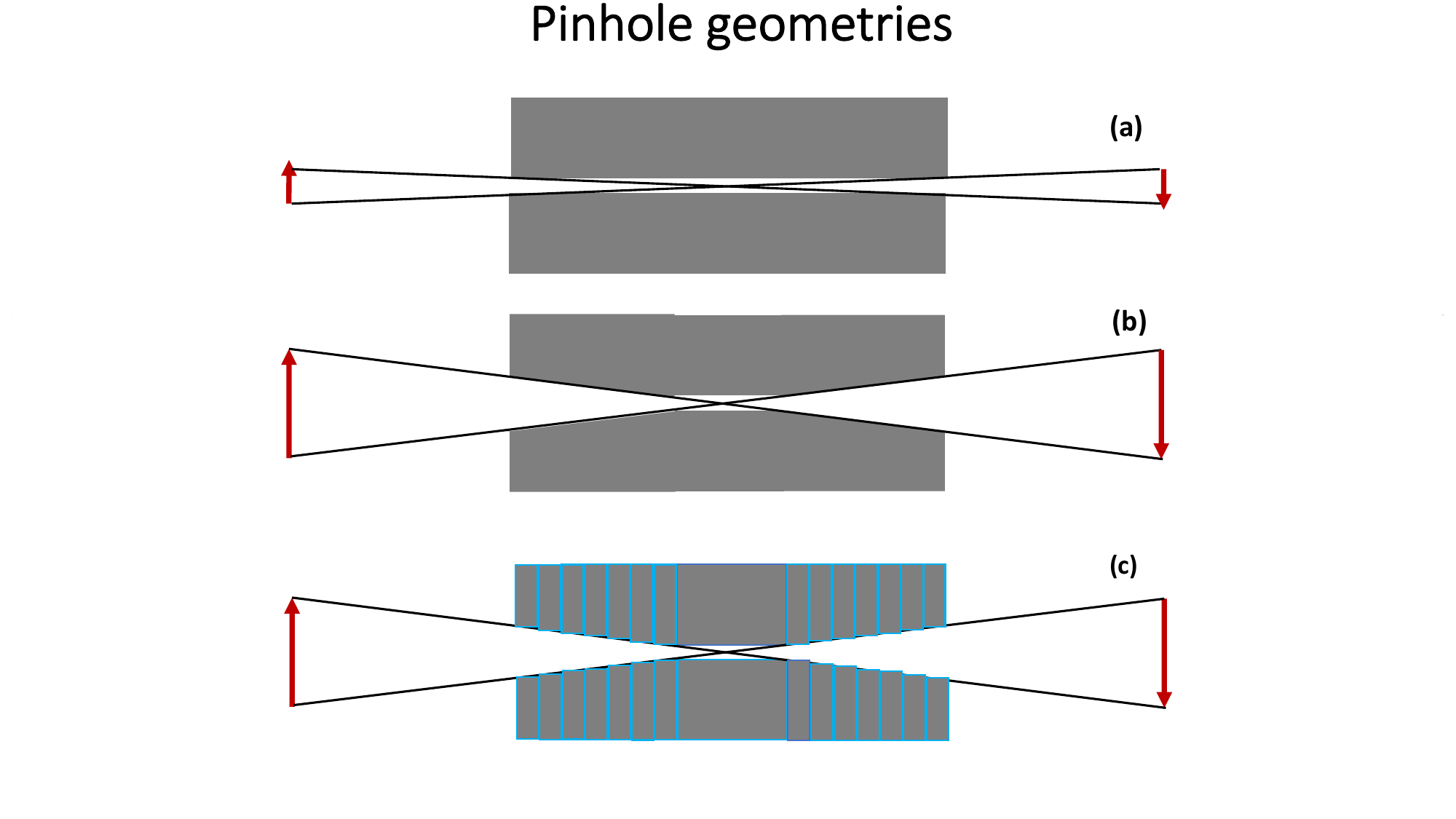}
\caption{ A set of possible pinhole configurations. (a) A cylinder with a simple center bore, which results in a small field of view, (b) a  straight cylinder with conical tapers on each side, (c) a  straight cylinder with  stepped collimators for the tapers.
}
\end{figure}
When  tapers are included, the attenuation of the neutrons is lower than for center bore cylinder of the same length, causing the PSF to be
wider and the spatial blur function to be wider. In addition, the total number of neutrons transmitted increases, so that the overall intensity of the image is higher.

For UGT diagnostics, the best pinhole had a center, cylindrical bore of approximately 0.33 mm diameter and  about  one neutron MFP long \cite{King}. 
A taper of tungsten alloy then extended the pinhole by approximately 30 cm before and after the central bore. 
Early on, the taper was constructed of a stack of approximately 1 inch thick cylinders (referred to as a stepped collimator), similar to Fig. 1c. 
In the 1980’s and 1990’s, the technology was developed to enable the tungsten to have smooth tapers without such steps. 

\section{The Point Spread Function for  realistic tapered pinholes}
In the present work, we wish to determine how much the size of an image can be changed by the  PSF for a realistic pinhole. 
For the size of the image, we use a metric known as $A80$. $A80$ is the area of the image that contains 80\% of the neutrons. 
Practically speaking, $A80$ can be calculated by ordering the image pixels (each of known area) from lowest to highest neutron intensity and 
summing the number of pixels needed to make up 80\% of the image area. We note that one could choose, for example, to consider the metrics $A50$ or $A90$, corresponding to
50\% or 90\% of the image area, and this choice does not change the main conclusion of our analysis discussed below.
 
To study the effect of the PSF on the area of the measured image, we calculated the PSF using different models.
These included (1) the Walton ray-tracing model that includes neutron absorption in 
the pinhole system but does not include neutron scattering, 
and  (2) a full MCNP \cite{MCNP} simulation that includes all neutron and gamma-producing interactions as the neutrons are transported through the line of sight.  
In addition, we compare full MCNP simulations for a 
conical tapered approximation with a 
stepped collimator geometry.
As a realistic pinhole example, we chose the stepped cylinder pinhole design for the Hermosa shot.
 
In Fig.2a, we show the PSF derived from a 3-D ray-tracing pinhole code that implements the Walton model, in which the stepped collimator is replaced with a smooth conical taper. As expected, the introduction of the tapers causes the PSF to go from a simple top-hat to a top-hat with broad wings.
Fig.2b compares the latter model with the full MCNP simulations for both the smooth taper approximation and for the full stepped collimator geometry.
We find that the gamma contribution to the image is small ( $<$0.1\%), and that 
the neutron scattered contribution is very low.
From this we conclude that the assumptions made in the Walton model are adequate and the results match both smooth conic tapes and the stepped cylinder pinhole
 MCNP analyses quite well, Fig. 2b. 
There is a distinct increase in the total neutron transmission with the stepped collimator geometry
compared to the tapered examples.
However, the overall normalization or  degree of neutron transmission  does not play a role in determining the area of the image. 
\begin{figure}
\includegraphics[width=4.0 in]{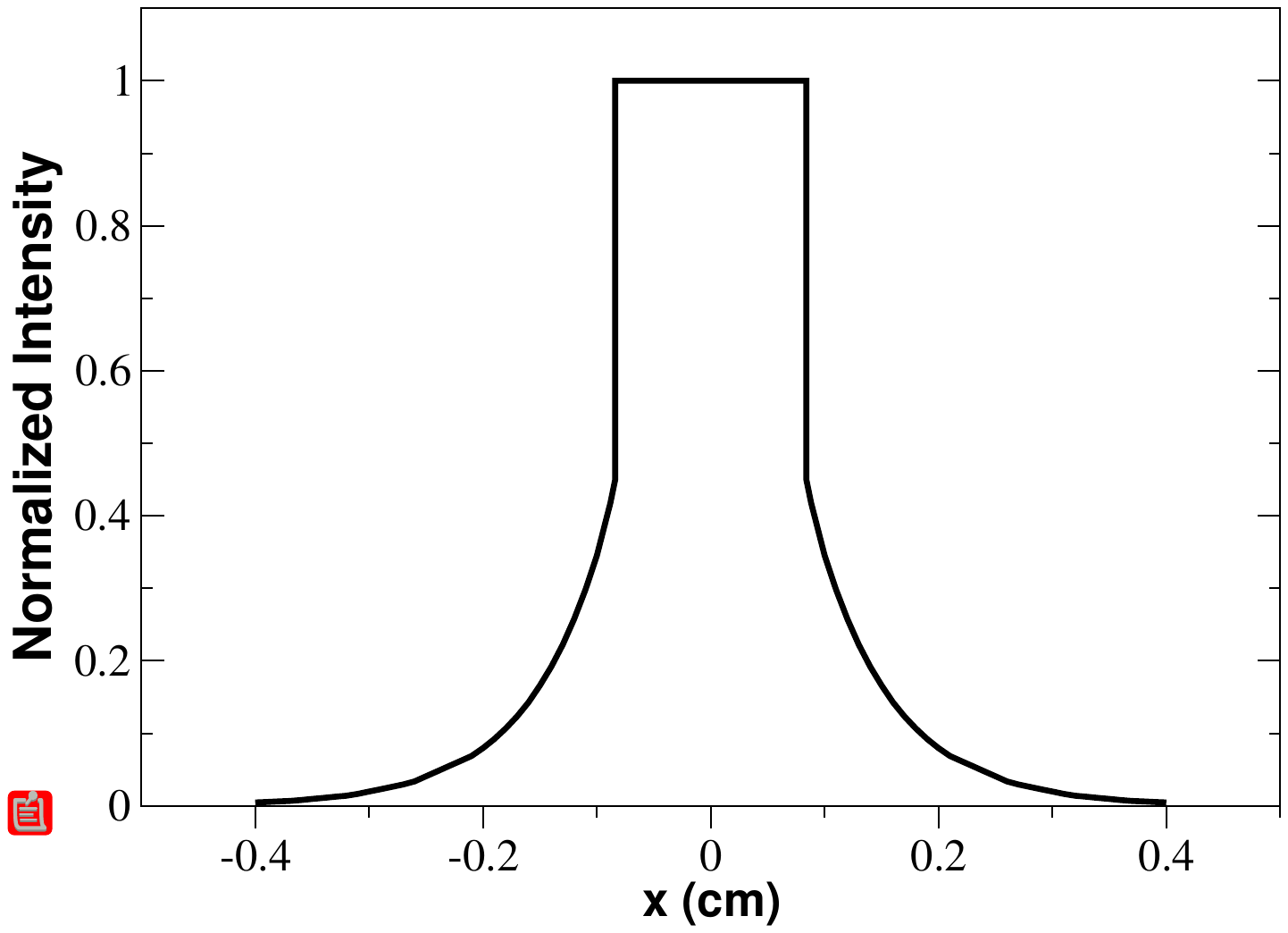}
\includegraphics[width=4.0 in]{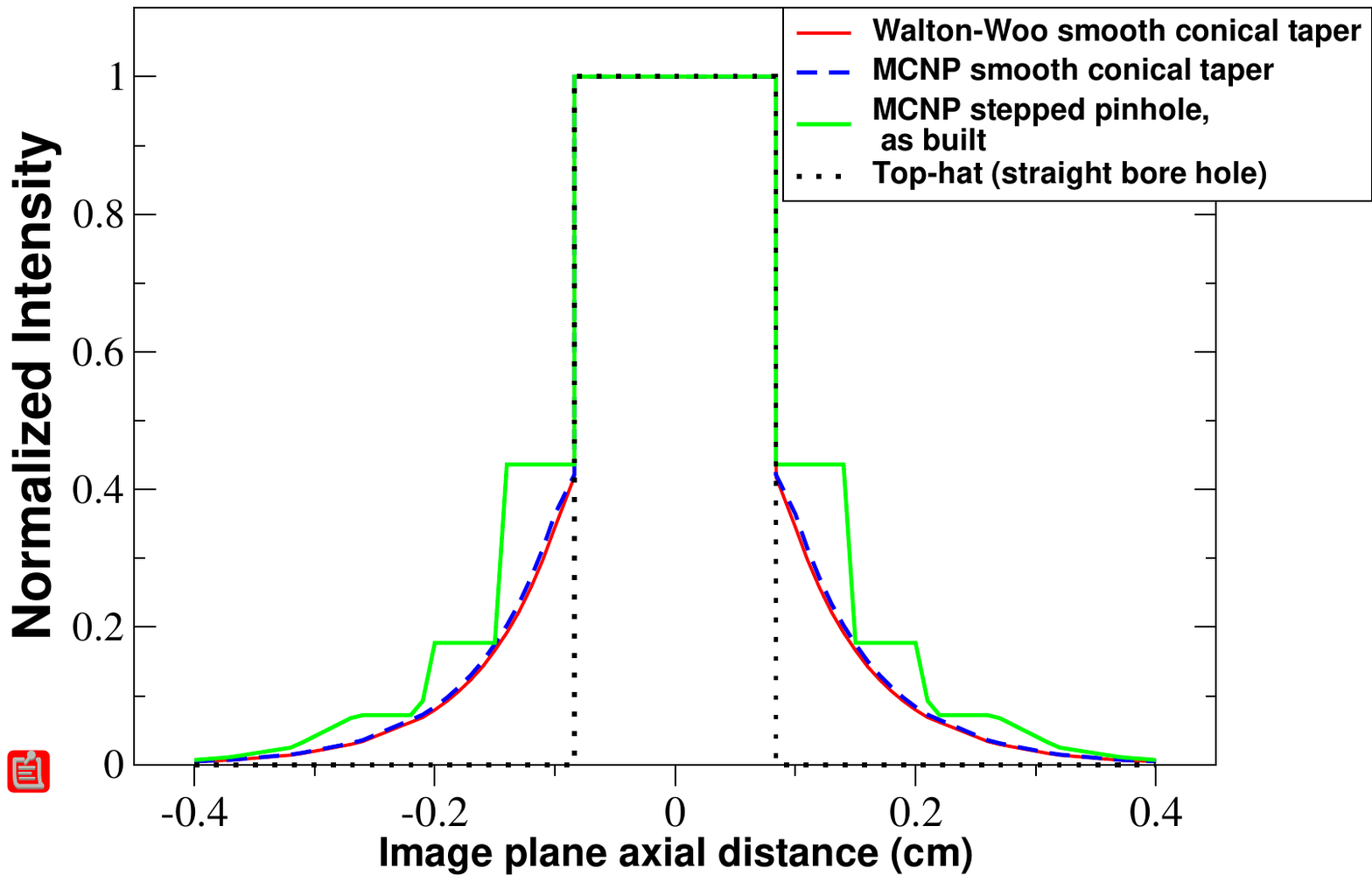}
\caption{The upper graph shows the PSF for the pinhole used in the Hermosa shot, as predicted by the Los Alamos Walton-Woo 3-D ray-tracing code, when a smooth conical taper is used to approximate the stepped collimator geometry that was actually used.
 The lower graph  compares the PSFs in the image plane derived from the MCNP model for the stepped pinhole with that for the  smooth conic tapers.
Also shown is the MCNP prediction for the smooth conic taper approximation and the top-hat PSF for an equivalent non-tapered straight bore hole pinhole.}
\end{figure}
\begin{figure}
\includegraphics[width=3.50 in]{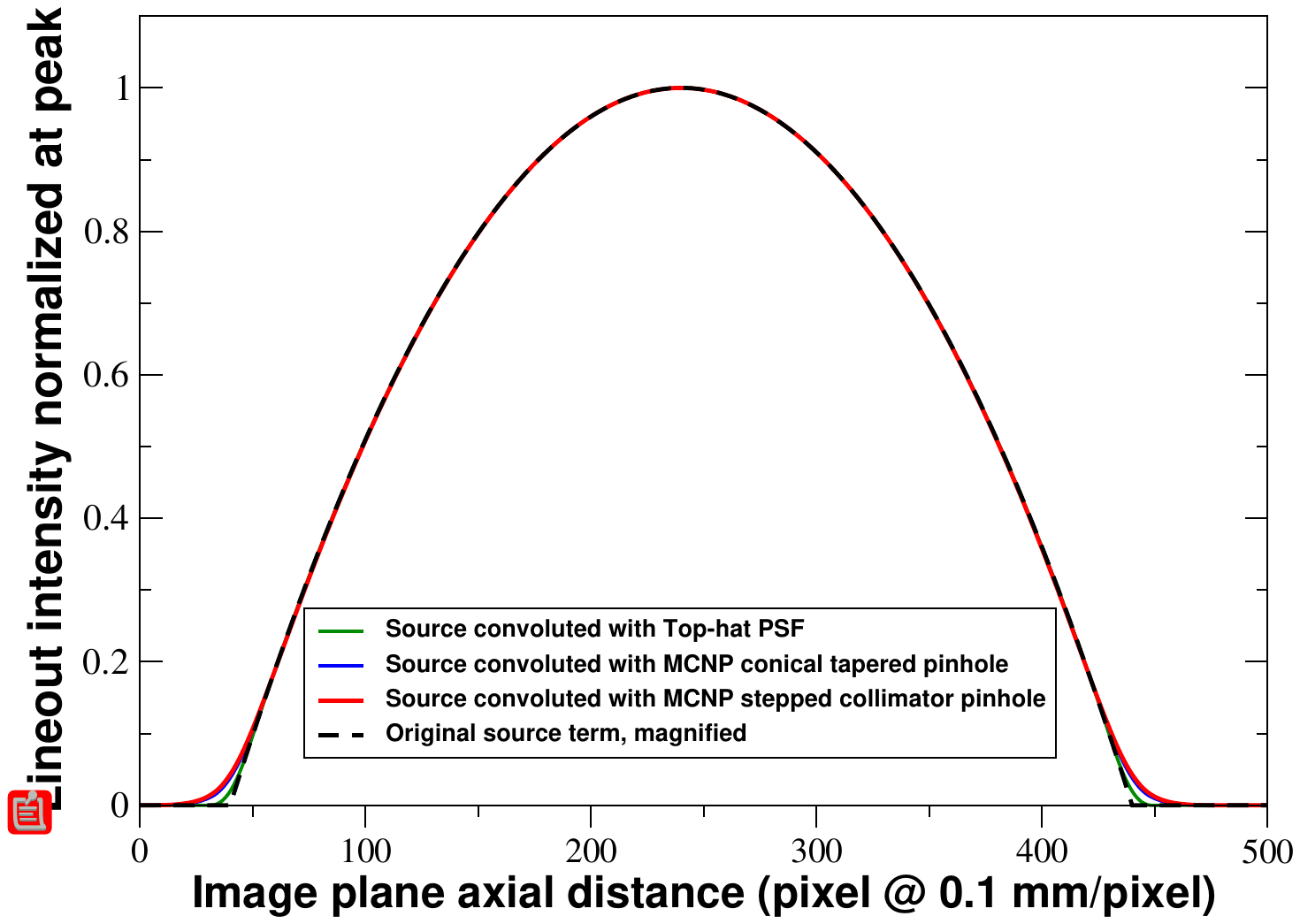}
\includegraphics[width=3.50 in]{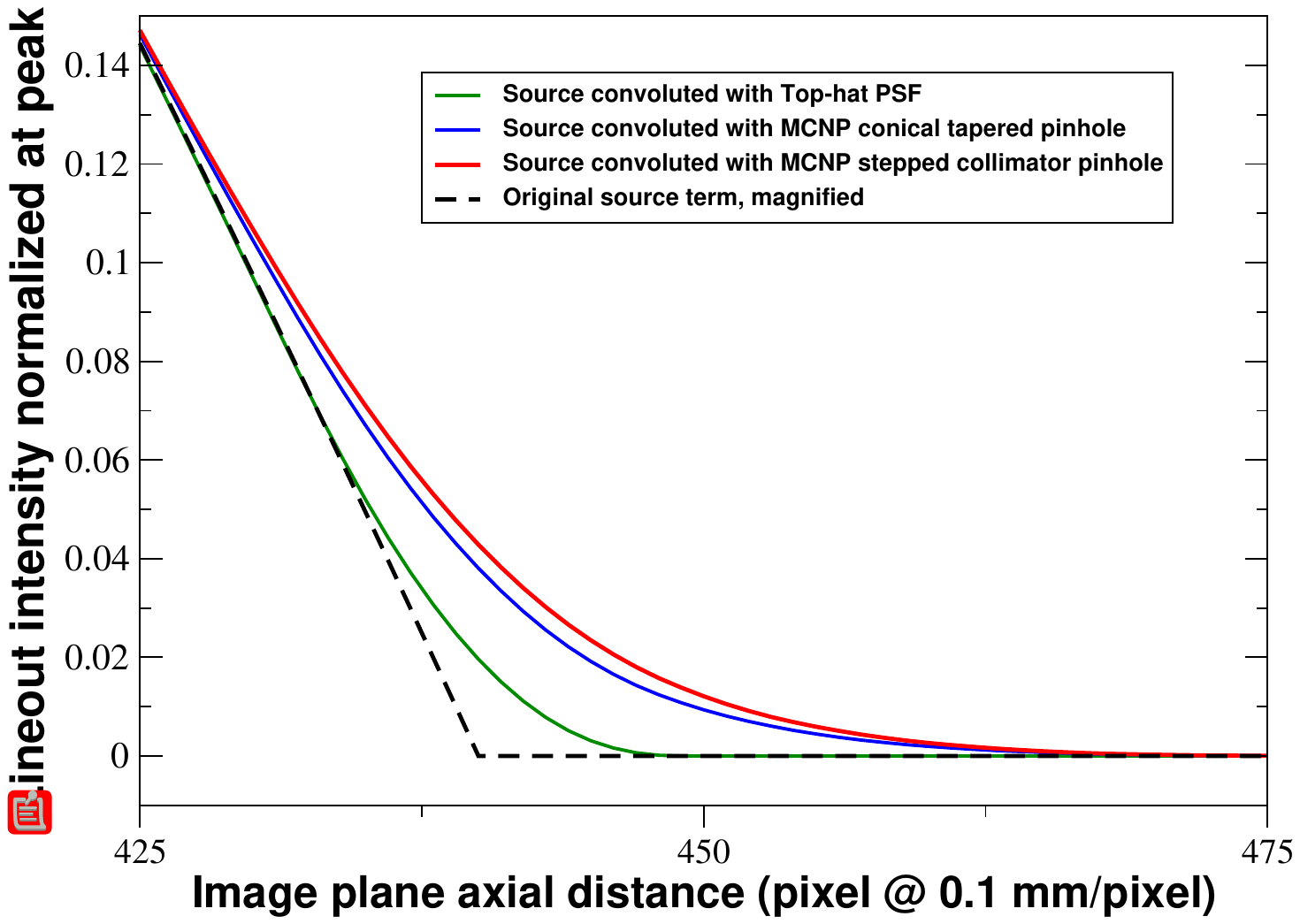}
\caption{Lineouts of the images blurred by the PSF compared to the lineout for original image.
(left) Lineout of the full images over 500 pixels. (right) The last 50 pixels of the same lineouts.
There is little difference seen between the blurred images and the original image and also between 
the different models for the pinhole, except at the outer edges of the images.}
\end{figure}

\section{The $A80$ area of the images}
To examine the effect of the pinhole geometry on image size we need to make an assumption about the nature of the  
 neutron source, and for this we use a very simple inverted parabola shape. 
To form images of the source in the plane of measurement, we perform a 2-D convolution of the magnified source
image with the different PSFs that are discussed above and compare these with the magnified source image in the same plane when the pinhole blurring is ignored.  
Fig. 3a and 3b show  lineouts from the resulting  blurred images in the image/measurement plane, compared to the original image in the same plane. 
As can be seen, there is little difference in the shape and size of the blurred images compared to the original, after
all images are renormalized.
In addition, the results for the tapered and stepped pinhole geometries are very similar. 

Pinhole PSFs can extend many pixels further than the center bore, so that the resulting convolution includes small but nonzero values far from the center bore. 
Typical PINEX images have background noise of the order of $\> 1\%$, which we represent as $\eta(x,y)$ in eq. 1.
Therefore, in PINEX analyses it is standard practice to threshold the image at about the $1\%$ level 
before performing the calculation of $A80$. 
After thresholding at the $1\%$ level, we find little difference between the $A80$ values for the different pinhole assumptions, Fig.4.
The same is true if we consider other area metrics, such as $A50$, $A90$ or any $AX$.
Therefore, we conclude that the finite size and blurring effects of the pinhole do not significantly
 affect the image size. 
For the simple inverted parabola toy model for the neutron source used here, 
the  maximum increase in size relative to the original magnified image occurs for the stepped collimator and corresponds to a  $\sim 1.5\%$ increase for $A80$.
The difference between a full MCNP treatment of the stepped collimator pinhole and a `perfect' top-hat cylindrical bore hole approximation for the pinhole is +0.8\%.
\begin{figure}
\includegraphics[width=3.5 in]{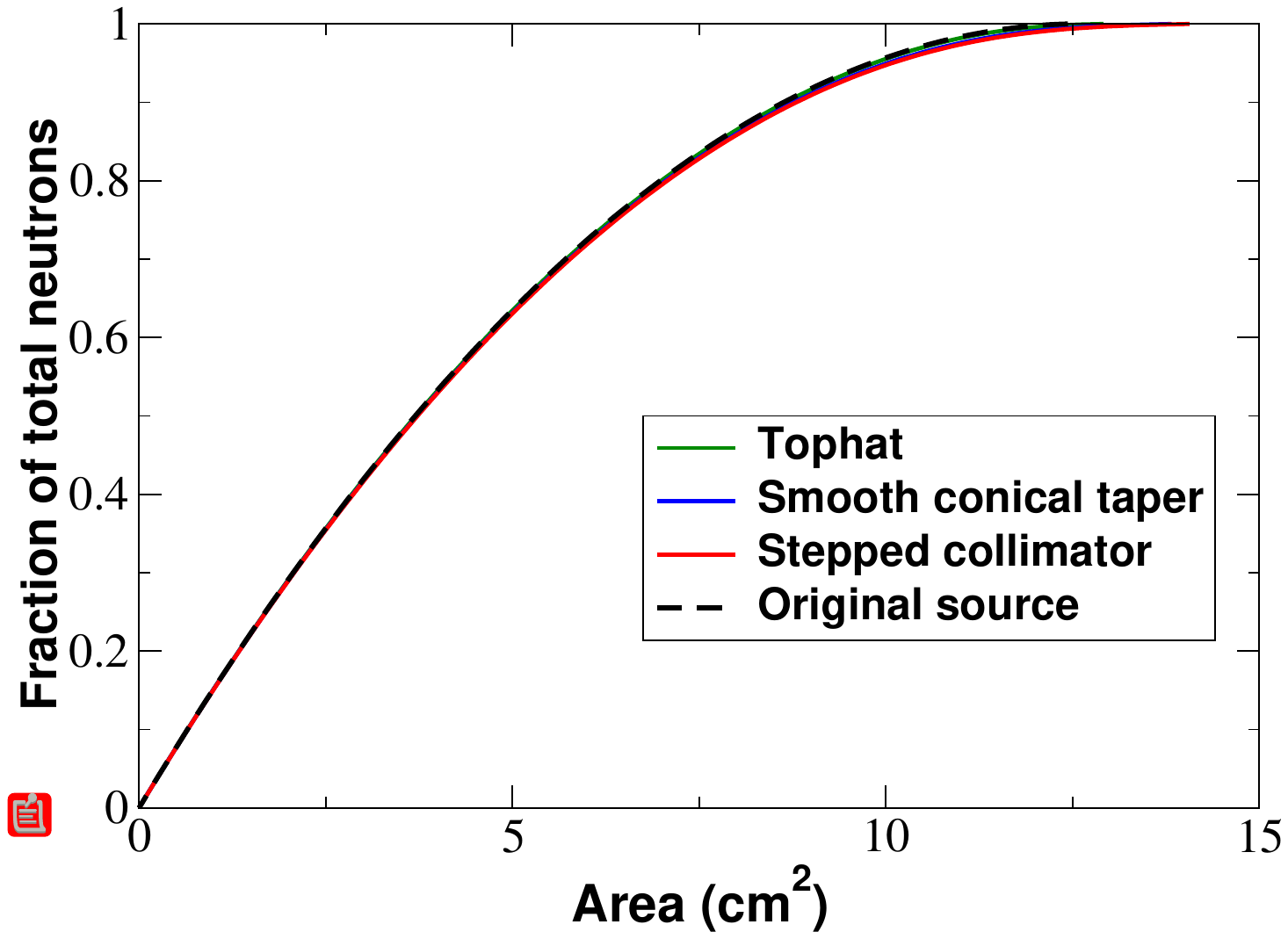}
\includegraphics[width=3.5 in]{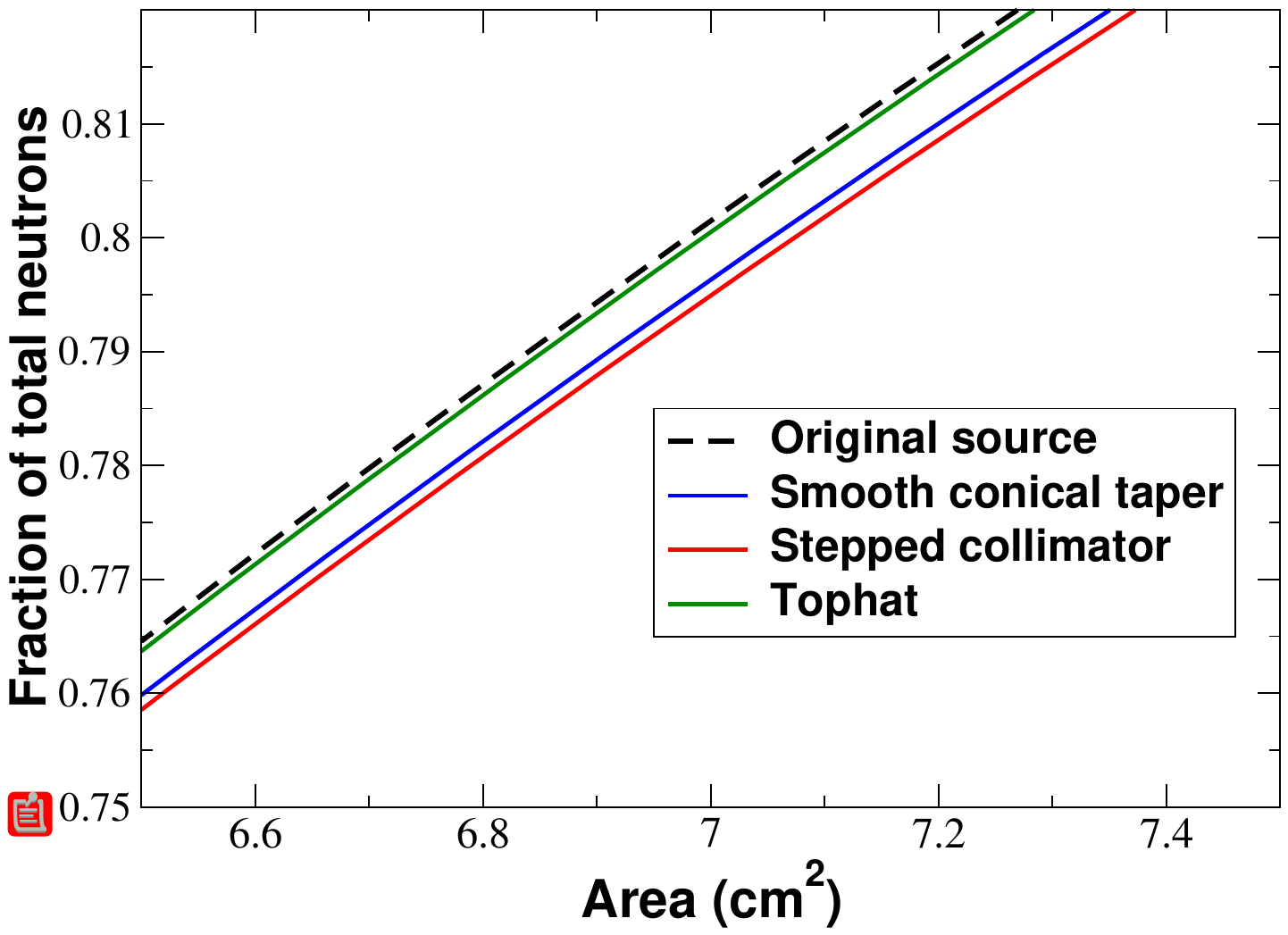}
\caption{The area of the image in the image plane as a function of the fraction of neutrons contributing to the image.
The pinhole PSFs extend many pixels further than the center bore and the resulting convolution includes small but nonzero values far from the center bore. For this reason, a threshold at the 1\% level was applied before calculating the areas associated
with neutron fractions. 
We then find that $A80$ value from a realistic MCNP treatment of the stepped collimator pinhole 
differs from the $A80$ value of the original image by at most 1.5\%. 
The difference between the former treatment and a `perfect' cylindrical bore hole (top-hat) treatment is +0.8\% }
\end{figure}
The precise magnitude of the increase found in $A80$ for the different treatments of the pinhole are dependent
on the shape of the neutron source shape assumed.
Shapes other than that assumed here could lead to larger increases in $A80$, 
and increases of the order of 5-8\% could be realistic.
Finally, we note that the lack of any significant effect of the pinhole on the size of the image can be understood intuitively from the large difference in the size of the source image relative to the size of the pinhole, Fig.5. In this 
picture all images have the same spatial scale and are in the same plane.
\begin{figure}
\includegraphics[width=4.0 in]{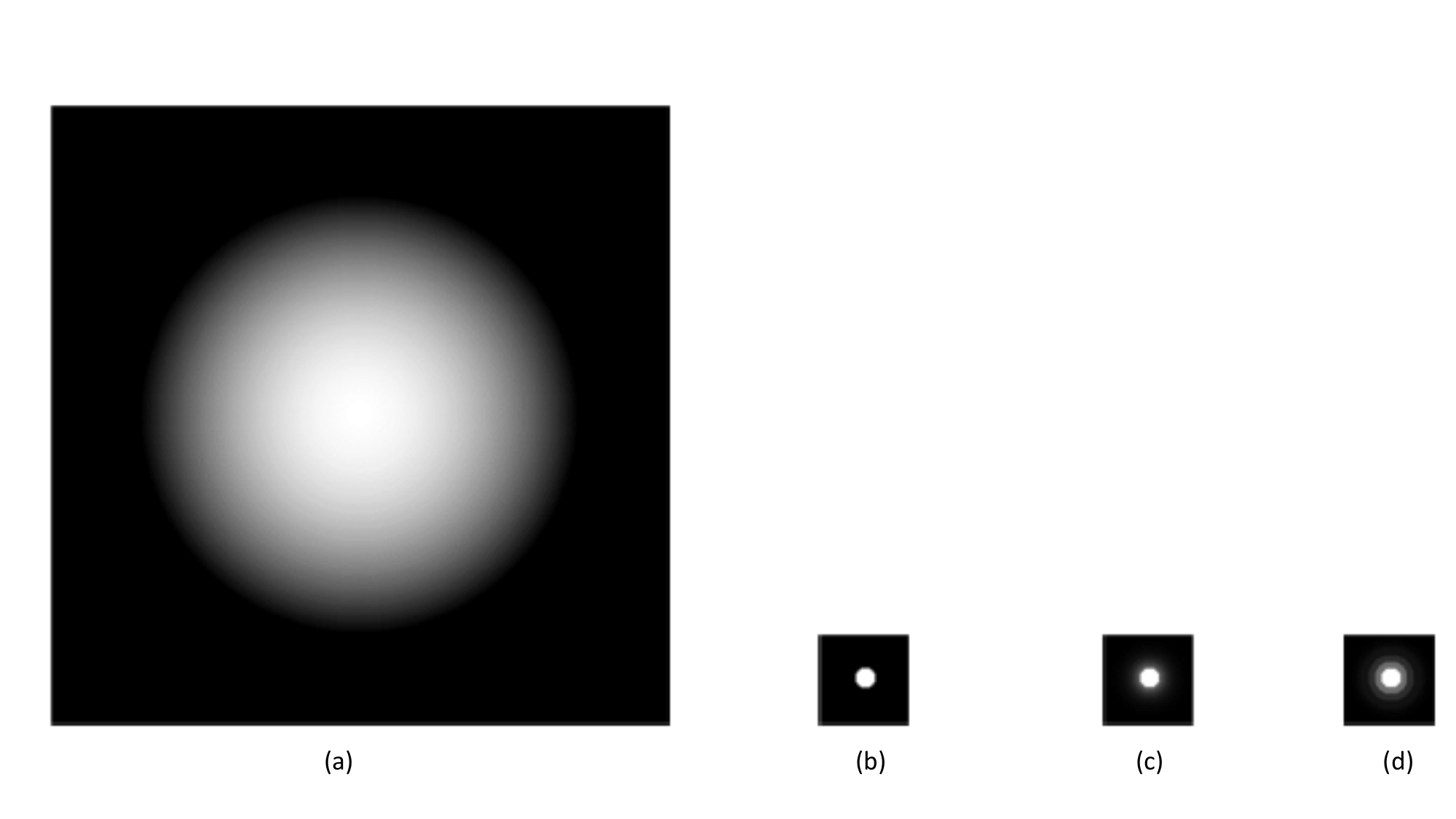}
\caption{Images of (a) the source term  and (b-d) the three pinhole geometry PSFs: (b) the `perfect’ tophat pinhole with no tapering, (c) the smooth conical taper pinhole, and (d) the stepped collimator pinhole.  All images are at the same spatial scale and in the same plane. The large difference in size between the source image and the pinhole 
geometries provides a qualitative explanation for why the pinholes do not change the image size significantly.}
\end{figure}
\section{Conclusions}
We examine the effects of pinhole blurring on the size of neutron images. For this, we chose a very simple
inverted parabola shape for the neutron source While different pinhole geometries allow
different degrees of neutron transmission to the image plane, 
they do not change the size of the image significantly. 
Using the area metric $A80$ for image size, and invoking a 1\% thresholding, 
we find that realistic pinhole geometries increases the metric by 1.5\%.
Larger increases in $A80$ ($\sim5\%$) from pinhole blurring would be possible with different assumptions for the shape of the 
neutron source.

\end{document}